\documentclass[12pt]{article}
\usepackage{fullpage}

\begin{document}

\title{Conferences vs. Journals: \\
Throwing the baby out with the bath water?}

\author{Claudio Gutierrez \\
Computer Science Department \\
 Universidad de Chile}
\date{March 2010}
\maketitle

 Criticism of the conference model should be put in context. 
Evidence suggests that the essential features of this model have emerged 
as responses to challenges posed by current trends of scientific research 
and the impact of the new techno-economic paradigm: the age of Information and Communication Technology~\cite{Freeman}. 
 The economic, social and technical impact of this revolution
 on the whole scientific research cycle is beginning to 
be assessed~\cite{Gray,Nowotny}.
 At the center of it is our discipline, computing science (CS),
that in this process is reflecting on its identity and
relationship with other disciplines~\cite{Denning,Shaw}.

 This context seems indispensable when  discussing today's problems
of scientific evaluation, in particular  the Conference vs. Journal 
(CvJ) debate. This debate, also, would benefit from systematic
historical and sociological studies of these practices. 
This ground would help to ponder the criticism of the conference model
by showing that a great percentage of the problems 
like the rapid increase in the number of submissions,
skimpy and slow reviews, declining paper quality,
pressures from academia and publishers, 
are affecting the whole system, be it conference or journal~\cite{Crowcroft,Smith}. In this article we briefly develop these arguments.

\paragraph{Evaluation in Classical Sciences.}
Although their origins can be traced  to scientific societies,
it is in the nineteen century that the journal and
peer review models as we know them today emerged.
The system evolved without major changes until the second 
half of the 20th century, when it was shaken
by the new printing and communication technologies (photocopies, email, 
Web, etc.).
Editorial boards evolved 
 from a group of people whose main goal was to find enough 
material to fill the pages of the journal, into a huge 
filtering system whose goal turned to 
enforce minimum standards and discard unsound works~\cite{Burnham}. 
Recognized today as the essence of scientific evaluation,
peer review has been lately the center of heated discussions
in all disciplines~\cite{AMS,Haack,Smith}. 
The journal model, in its origin had two driving forces:
 the need to incorporate more articles into the publishing process,
and  to encourage scientists to disclose their work
(by giving credit and prestige to the author)~\cite{merton}.
The first one is not needed anymore, but a new one has emerged: 
to measure ``productivity'' in a system of science where 
accountability plays an increasing role~\cite{Nowotny}.

  Regarding our CvJ debate, another facet of this history seems more relevant.
Scientific societies held periodic meetings where their members
presented their works and advancements in their disciplines.
At the beginning, journal contents were almost a compilation 
of these presentations. The method did not scale
and these practices split: on one hand today's journals,
and on the other, periodic meetings of the society
 where members used to present their findings and ideas.  
Journals  got under
evaluation methods to check the relevance of submissions,
while meetings kept a natural filter: to be member of the society.
 These two forms of communicating scientific research made their
own way.

\paragraph{The advent of Conferences.}
  For some sociological reason, a strange event,
a mix of meetings and journals, was developed in our discipline.
Opinions and remembrances
ascribe this phenomena to the particular
nature of the ``products'' of the discipline; 
the perception that CS has a different speed from other sciences;
 the few journals available and
the long delays in refereeing and publishing;
 facilities given by traveling in the jet age; 
 the avoidance of wasting time in polishing for a journal 
material already presented in conferences.
(See CACM, vol 52, Nos. 1,4,5,8 and links from there).
  The need to test these hypotheses becomes evident when 
reading the most widespread of them:
  ``Quick  development of the field required quick review and 
distribution of results'', which although intuitive, does not explain
conferences ({\em Letters} in other fields
give  much ``faster'' review and dissemination times). 
    The phase transitions of this process are also important.
Table~\ref{acm}, showing rough statistics
based on the information of the ACM Digital Library, 
indicates that conferences/workshops are
influential in the field at least since the early 1970's when
the numbers exploded.
The series LNCS,  which publishes mainly CS events proceedings, 
began in the seventies with a dozen a year and since then 
it has grown  to reach currently more than five hundred a year. 
 The data, although
illustrative, is by any means apt to derive definitive conclusions, 
because parameters like attendance, impact, spreading, relevance, etc. of 
these events is not weighed. 
Definitively a systematic study is needed to put the CvJ 
debate on solid grounds.

\begin{table}
\begin{center}
\begin{tabular}{|l|c|c|c|} \hline
Period & Journals & Conf. (still existing) & Conf. (not existing today) \\ \hline
Before 1960 & 1 & 0 & 2 \\
1960-64 & 0 & 2 & 0 \\
 1965-69 & 1 & 7 & 0 \\
 1970-74 & 0 & 11 & 3 \\
 1975-79 & 4 & 4 & 2 \\
 1980-84 & 3 & 8 & 2 \\
 1985-89 & 0 & 23 & 14 \\
1990-94 & 4 & 22 & 18 \\ 
1995-99 & 3 & 46 & 13 \\
 2000-04 & 9 & 55 & 25 \\ 
 2005-09 & 13 & 38 & 6 \\ \hline
\end{tabular}
\end{center}
\caption{Number of ACM Journals and Conferences by birth date}
\label{acm}
\end{table}

\paragraph{Conferences vs. Journals.}

  A big part of the discussions on CvJ focuses on criticism on the
current conference model.
Bad quality of papers, splintering of  communities,
 few attendees, reviews done under extreme time and workload pressures,
are among the typical complaints. 
  Many of them are also found in current criticism of journals.
 Thus to contribute to the debate, it seems important to
isolate characteristics that are inherent to conferences.
 Below we present a first approximation.

\begin{itemize}

\item {\em Time constraints for the researcher}. 
Enforcing research {\em deadlines}  puts time pressure.
  On the contrary, journals allow submissions at any moment,
giving more freedom and relaxed pace to research.
 Deadlines are slightly being incorporated into journals,
e.g. ``special issues'', Proc. of the VLDB Endowment, etc. 
Regarding time constraints {\em for  reviewers}, the conference practice
is being increasingly adopted by journals, by fixing review times.

\item {\em Space constraints}. Limiting the size of articles
 puts pressure on the form of communication.
  Only the essential without details is presented.
The motivations seem to be time review constraints
and fast communication.
 This notion is rather foreign to classical journals, although
 {\em Letters} have incorporated it long ago.

\item {\em Boolean decisions}. This is one of the distinctive characteristics
of the conference model. As the number of submissions grow,
 many journals are incorporating sharper decisions methods 
(limited resubmission rounds, etc.)
Under the think/act dichotomy, this method incentives 
action rather than deliberation.

\item {\em Discussion post-publication}. Real testing of hypothesis
begins after the conference (where only a sketch  --``extended
abstract-- is expected) under an open-world scrutiny.
On the contrary, journals expect articles to be definitive 
once published. Testing is done before by reviewers in a closed world.

\item {\em Evaluation by batch.} Evaluation is done on a set of submissions
at a time. Hence, implicitly the value of a paper is weighed against
their peers in the same event. 
This method introduces the notion of ranking among submissions 
and  bias in favor of fashions, groups, and mainstream developments.

\item {\em Changing board of editors}. 
The notion of  {\em Program Committee} (PC) 
is in my opinion one of the cornerstones of the system. 
Complemented by the  Steering Committee ensuring long term 
stability, PC's reflect and enforce the mobility of areas, topics, people
and criteria. 
  Today
 journal editorial boards are rather closed and stable groups, 
not without inbreeding, limiting innovation.

\item {\em Dissemination of ideas done by the producer}.
 I think this is the kernel of the conference model, and a feature 
not sufficiently highlighted in the discussion on CvJ.
 Conferences/workshops are (easily) organized by the interested
researchers  in a new field, who  initiate a spiral 
 of contributions/evaluation/dissemination that
ultimately will shape a new area if persisting in time.
 The key point is that there is no need of intermediaries, and the
growth process parallels  that of the community. In this sense,
it resembles the origins of scientific societies that shaped
current branches of knowledge.
(There is a significant link between this feature 
and current discussions on the future of publishing, open journals, etc.)

\item {\em Physical presence, face-to-face meeting}.
  This has been assumed to be the most distinctive characteristic
of the conference model.
What is surprising and intriguing is that the community 
leading virtual communications and immaterial production 
is the one that seems more prone to physical meetings. 
Is this a contradiction? 
Let us remark that in our age of rapid communications, 
the other  characteristics
of conferences discussed do not need  physical contact or presence.

Being aware  of the role and power of
--and in favor of enforcing--  face-to-face communication
and physical bonds, I rather disagree that it is the essence 
of what has rocketed the conference model. 

\end{itemize}

\paragraph{Final Comments.}

The conference model  reflects fairly well 
our times. Like it or not,
the first three characteristics discussed are typical of 
today's social and economical rhythms,
and fit very well modern time constraints in any area.
Arguments stating that the conference
 ``has fractured the discipline and skewered it
towards short-term deadline driven research'' point rightly 
to consequences, not to causes.
 In fact, these are general trends observable in 
all disciplines. 
Many of the aspects blamed to conferences are
part of a new paradigm of science development
(called ``Mode 2'' in the literature)
characterized by the steering of research priorities, accountability
of science and commercialization of research. 
    This mode of development --their advocates state--, 
 would have superseded the classical model (``Mode 1'')
 characterized by the hegemony 
of theoretical or, at any rate, experimental science, 
by an internally-driven taxonomy of disciplines,
and by the autonomy of scientists and their host institutions, the 
universities~\cite{Nowotny}. 

  Definitively, when reading the CvJ debate, one cannot but
bring to mind broader discussions on the  transformation of the 
scientific method in our age.
 I think we should incorporate them into our discussion 
in order to fairly assess the contributions of conferences 
to current scientific developments.

\end{document}